\documentclass[preprint,12pt]{elsarticle}

\usepackage{amssymb}
\pdfoutput=1

\journal{Chemical Physics Letters}

\def\cm-1{cm$^{-1}$}

\def\MolOne{\textbf{1}}
\def\MolTwo{\textbf{2}}

\begin{document}

\begin{frontmatter}

\title{Direction-dependent secondary bonds and their stepwise melting in a uracil-based molecular crystal studied by infrared spectroscopy and theoretical modeling}

\author[Wigner Research Centre for Physics]{Zsolt Szekr\'enyes}
\address[Wigner Research Centre for Physics]
{Institute for Solid State Physics and Optics, Wigner Research Centre for Physics, Hungarian Academy of Sciences, H-1525 Budapest, Hungary}
\author[BME Lendulet Quantum Chemistry]{P\'eter R. Nagy\corref{cor1}}
\ead{nagyrpeter@mail.bme.hu}
\address[BME Lendulet Quantum Chemistry]{MTA-BME Lend\"ulet Quantum Chemistry Research Group, Department of Physical Chemistry and Materials Science, Budapest University of Technology and Economics, P.O. Box 91, H-1521 Budapest, Hungary}
\author[ELTE Molecular Spectroscopy]{Gy\"orgy Tarczay}
\address[ELTE Molecular Spectroscopy]{Laboratory of Molecular Spectroscopy, Institute of Chemistry, E{\"o}tv{\"o}s Lor{\'a}nd University, H-1518 Budapest, Hungary}
\author[University of Namur (FUNDP)]{Laura Maggini}
\author[University of Namur (FUNDP),Cardiff University]{Davide Bonifazi}
\address[University of Namur (FUNDP)]{University of Namur (FUNDP), Department of Chemistry, Rue de Bruxelles 61, 5000 Namur, Belgium}
\address[Cardiff University]{Cardiff University, School of Chemistry, Park Place, CF10 3AT Cardiff, United Kingdom}
\author[Wigner Research Centre for Physics]{Katalin Kamar\'as\corref{cor1}}
\cortext[cor1]{Corresponding authors}
\ead{kamaras.katalin@wigner.mta.hu}

\begin{abstract}

Three types of supramolecular interactions are identified in the three crystallographic directions in crystals of 1,4-bis[(1-hexylurac-6-yl)ethynyl]benzene, a uracil-based molecule with a linear backbone. These three interactions, characterized by their strongest component, are: intermolecular double H-bonds along the molecular axis, London dispersion interaction of hexyl chains connecting these linear assemblies, and $\pi$--$\pi$ stacking of the aromatic rings perpendicular to the molecular planes. On heating, two transitions happen, disordering of hexyl chains at 473 K, followed by H-bond melting at 534 K. The nature of the bonds and transitions was established by matrix-isolation and temperature-dependent infrared spectroscopy and supported by theoretical computations.

\end{abstract}


\end{frontmatter}

\section{Introduction}
\label{sect:intro}

H-bonds formed in uracil-based molecules have played a major role not only in nucleic acid research\cite{Watson53, Saenger84,Jeffrey91,Jeffrey97}, but recently also as the secondary forces holding together two-dimensional networks with long-range order \cite{Piot09,Barth05,Surin07}.
Numerous studies on the role of H-bonding in uracil
and uracil derivative molecules have been reported \cite{Piot09,Troitino06,Hobza99,Llanes08,Feibush86, Lehn90,Palma09,Llanes11,Dordevic15}. The study of uracil derivatives is of special interest as these compounds are excellent models to demonstrate the key role of supramolecular chemistry both in biochemistry and materials science, such as the formation of two-dimensional porous networks on surfaces \cite{Piot09,Llanes08,Palma09,Palma08,Llanes09, Bonifazi09,Mohnani10,Marangoni13}.

\begin{figure}
\includegraphics[width=12 cm]{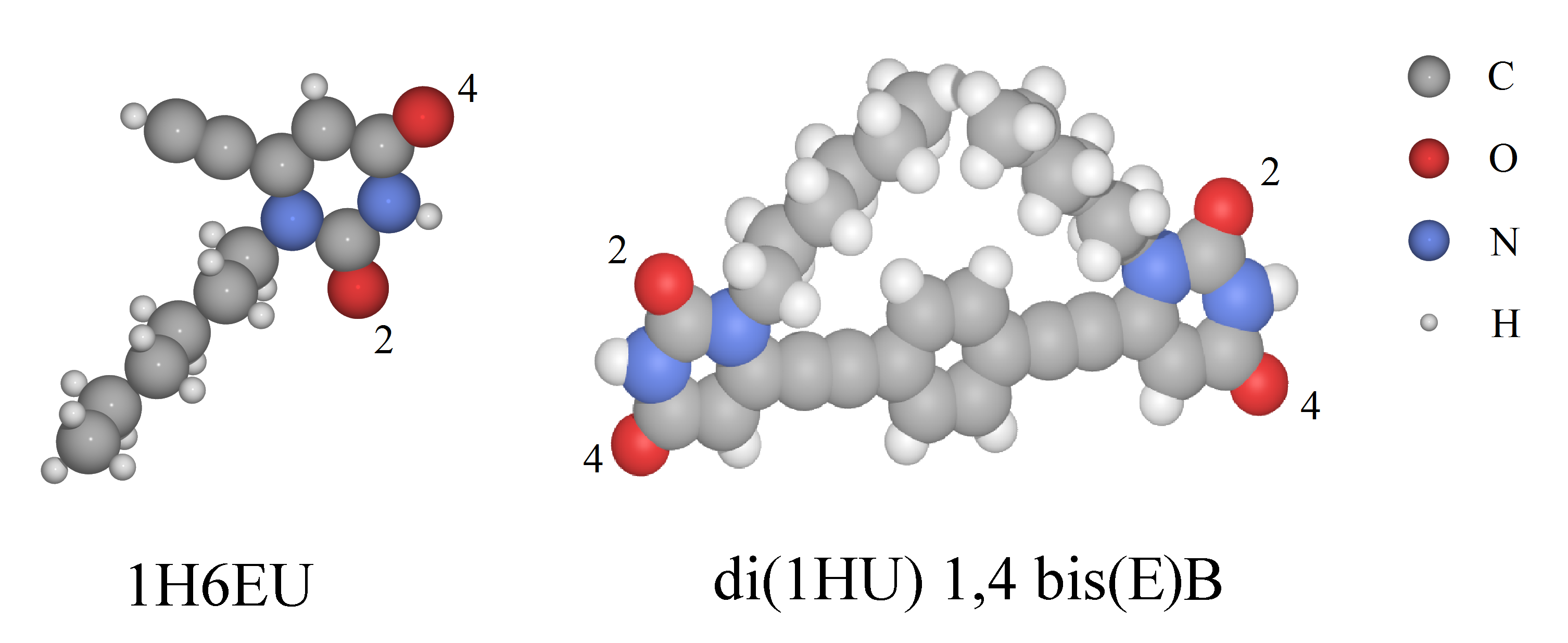}\\
\caption{{\MolOne} (a) and {\MolTwo} (b) have identical hydrogen bond forming functional groups (the carbonyl and amine groups). The two carbonyl groups numbered 2 and 4, respectively, may participate in different H-bonding motifs.}\label{fig:1}
\end{figure}

We present a study on a uracil-based molecular solid, formed of 1,4-bis[(1-hexylurac-6-yl)ethynyl]benzene (\textbf{2}), investigating the nature of
the interaction between molecular units in the crystalline environment by infrared (IR) spectroscopy.
Theoretical modeling on small oligomers was also utilized to assist the
assignment and interpretation of the measurements.
We have performed this study as a continuation of our work on a simple monouracil derivative (1-hexyl-6-ethynyluracil (1H6EU, \textbf{1}) \cite{Szekrenyes12}.  Two 1H6EU units (Fig. \ref{fig:1} (a)), connected by a benzene ring, form \textbf{2} (Fig. \ref{fig:1} (b)). Molecule {\MolOne} is a monotopic molecule with  H-bond-forming sites at one end, a prospective ingredient of dimer structures, while \textbf{2} is a ditopic module enabling the formation of a linear backbone as H-bond-forming functional groups are present at its opposite ends. This ditopic uracil derivative unit is a very good candidate for a linear linker in self-assembled supramolecular networks. {\MolTwo} has already proved its potential in forming 2D bicomponent hexagonal porous structures at the solid-liquid interface \cite{Palma08} as well as in molecular recognition through triple H-bonds \cite{Llanes08}.

In derivative {\MolOne}, we studied the type of H-bonds by comparing the matrix-isolation infrared spectra of individual molecules with those of the molecular solid; in addition, we followed the melting process of the H-bonds at elevated temperature by both experimental and theoretical methods \cite{Szekrenyes12}. Here we report a similar investigation with the result that just below the H-bond melting temperature, a different phase transition occurs in the solid, from a planar tetrameric to a stacked columnar structure. These two consecutive transitions indicate the relative importance of the $\pi$--$\pi$ stacking, H-bonding, and hexyl--hexyl dispersion dominated interactions.

\section{Materials and methods}
\label{sect:exp}

\subsection{Materials and experimental methods}

Molecules {\MolOne} and {\MolTwo} (Fig. \ref{fig:1}) were prepared according to previously published procedures \cite{Llanes08,Llanes09}. Temperature-dependent
infrared spectra in the solid state were recorded on powders ground in KBr pellets. Mixing with KBr dilutes the material so it becomes transparent to IR light, but the grain size in the pellets ($\sim 1 \mu m^3$) ensures that they still can be regarded as solids, preserving the structure.

The matrix isolation (MI) setup is described in detail
elsewhere \cite{Pohl07,Gobi10}. Briefly, the evaporated sample was mixed with argon (Messer, 99.9997\%) before deposition onto an 8-10 K CsI window. The gas flow was kept at 0.07 mmol min$^{-1}$, while the evaporation temperature was optimized to 353$\pm$5 K and 530$\pm$5 K for molecules {\MolOne} and {\MolTwo}, respectively. Under these conditions, the sample consists predominantly of isolated molecules, with only a small amount of aggregated species present. Therefore, comparing MI-IR spectra with solid-state KBr pellet spectra allows us
to study the effect of aggregation.

MI-IR spectra were recorded with a Bruker IFS 55 Fourier transform infrared (FTIR) spectrometer with 1 cm$^{-1}$ resolution and a deuterated triglycin sulfate (DTGS) detector, and the temperature dependence in KBr pellets with a Bruker IFS66v vacuum FTIR instrument with 2 cm$^{-1}$ resolution and a mercury cadmium telluride (MCT) detector. All spectra were taken in the 400--4000~\cm-1 range with a Ge/KBr beamsplitter. The baseline was corrected by an adjusted polynomial function.

\subsection{Theoretical methods}

Stabilization energies of noncovalent complexes (i.e. dimerization, trimerization, etc. energies) were computed by density functional theory (DFT). Equilibrium structures were optimized using the $\omega$B97X-D functional developed by Chai and Head-Gordon \cite{Chai08a,Chai08b} and the 6-311G(d,p) basis set. The $\omega$B97X-D is a long-range corrected
hybrid density functional with dispersion corrections and found to be reliable to describe various noncovalent interactions of
molecules of main group elements \cite{Riley10,Burns11,Goerigk11,Vydrov12}.

For each optimized structure, an additional single-point energy computation was performed with the $\omega$B97X-D/6-311++G(3df,3pd) method.
All reported stabilization energies are obtained with this protocol, except for the case of the van der Waals tetramer structure. This tetramer illustrates the structure of two neighboring unit cells in our hypothetical solid state structure introduced in more detail in Section \ref{sect:results}. However, without the surrounding crystal field, this tetramer geometry does not correspond to a local minimum with respect to the nuclear coordinates. To obtain an approximate substructure characteristic of the three-dimensional lattice, a constrained geometry optimization was performed. The tetramer structure was kept approximately in its plane by applying harmonic constraints on the torsion angles that set the planes of the eight uracil rings parallel to this plane. The obtained structure cannot be considered as a local minimum and accordingly, we do not deduce quantitative information from this result.

Let us note that since the present systems contain several noncovalent interactions of different nature,
it is important to take the basis set superposition error (BSSE) into consideration. The BSSE can be reduced significantly using a close to complete basis set, such as 6-311++G(3df,3pd). Intermolecular BSSEs were estimated for dimers of {\MolOne} and {\MolTwo} by computing counterpoise corrections \cite{Boys70}, were found to be smaller than 1~kcal$\cdot$mol$^{-1}$ in all cases and consequently have been neglected. All computations were performed by the Gaussian 09 program package \cite{G09prog}.

In the previous study on {\MolOne} systems, temperature-dependent ab initio molecular dynamics (AIMD) computations were performed in the canonical ensemble \cite{Szekrenyes12}.
These results were in good agreement with the experimental spectra and provided important aid to the assignment
and understanding of temperature-dependent IR experiments on {\MolOne}. The large size of the {\MolTwo} dimers-tetramers prevents us from computing free energies and infrared spectra for these systems, but due to structural similarities, considerations formulated for {\MolOne} based on AIMD data may
be transferable to {\MolTwo}. For instance, based on the same arguments as presented in Ref. \cite{Szekrenyes12}, only diketo tautomers were considered in case of all {\MolTwo} based systems.

\section{Results and discussion}
\label{sect:results}

\begin{figure}[h!]
  \includegraphics[width=15 cm]{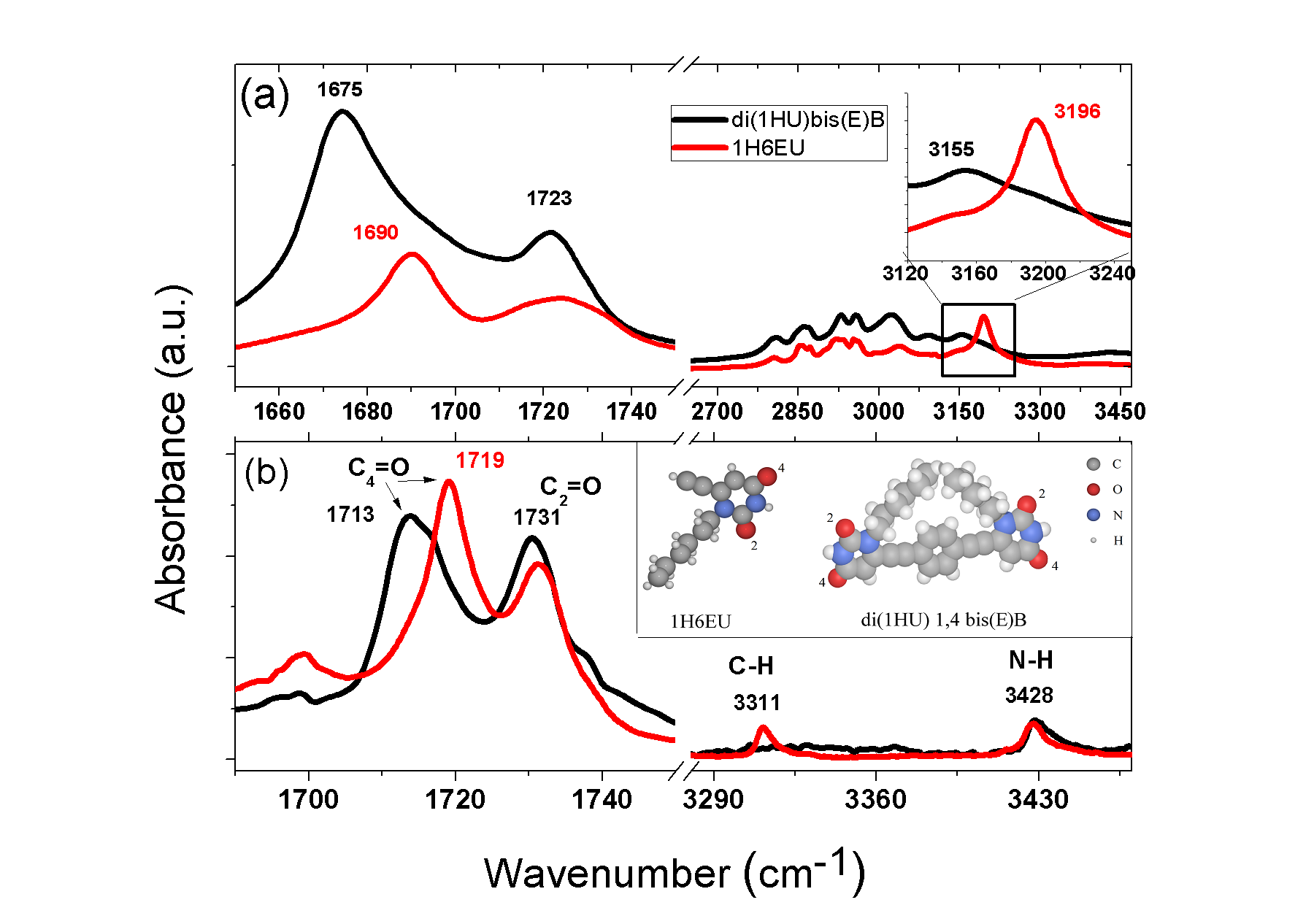}\\
  \caption{(a) Experimental room-temperature spectra of {\MolOne} and {\MolTwo} and (b) MI-IR spectra of the isolated molecular units. Subscripts 2 and 4 refer to the corresponding carbonyl groups according to Fig. \ref{fig:1}. The spectrum for {\MolOne} was taken from Ref. \cite{Szekrenyes12}.
  }\label{fig:2}
\end{figure}

Figure \ref{fig:2} presents the IR spectra of {\MolOne} and
{\MolTwo} recorded in the solid state at room temperature and
MI-IR spectra of isolated monomers recorded in an argon matrix at 8~K. We previously assigned the IR modes of {\MolOne} considering the solid sample as a static assembly of dimers with H-bonds constituting the main
intermolecular interaction \cite{Szekrenyes12}. Evaluating the possible dimer conformers of {\MolOne}, we found that the N-H (3) and one of the C=O groups (2 or 4) is always involved in H-bond formation, while the other C=O group is free. The band at 1690~{\cm-1} was assigned to the H-bonded carbonyl group while the band centered at 1723~{\cm-1} represents the free C=O stretching vibration. The band at 3196 {\cm-1} corresponds to the H-bonded N-H stretch mode. In analogy, for derivative {\MolTwo} we assign the H-bonded C=O and N-H stretch modes to the bands at 1675~{\cm-1} and 3155~{\cm-1} and the free C=O to that at 1723~{\cm-1}. While the free carbonyl mode is at the same frequency for molecules {\MolOne} and
{\MolTwo}, there is a considerable difference for the H-bond affected IR bands. These bands shift to lower wavenumbers for {\MolTwo} suggesting a stronger interaction of the molecular modules through H-bonds. In the
MI-IR spectra (Fig. \ref{fig:2} (b)) the mode assignment of the isolated monomers of derivative {\MolOne} was done by comparing with the theoretical spectrum computed for monomers at 5~K. The bands at 1719~{\cm-1} and at 1731 cm$^{-1}$ belong to the free C$_4$=O and C$_2$=O (according to Fig. \ref{fig:1}) vibrations of the monomer, respectively. For molecule {\MolTwo} the free
vibration of C$_4$=O and C$_2$=O is assigned to the band at 1713~cm$^{-1}$ and at 1731~cm$^{-1}$. The 6~cm$^{-1}$ difference between the C$_4$=O vibrational frequencies of the two molecules can be attributed to their structural differences. The band situated at 3428~{\cm-1} is assigned to the free N-H vibration and is at the same frequency for both units. Finally, the band at 3311~{\cm-1}, present only in {\MolOne}, corresponds to the ethynyl stretch mode $\nu(\equiv$C-H).

\begin{figure}
  \includegraphics[width=12 cm]{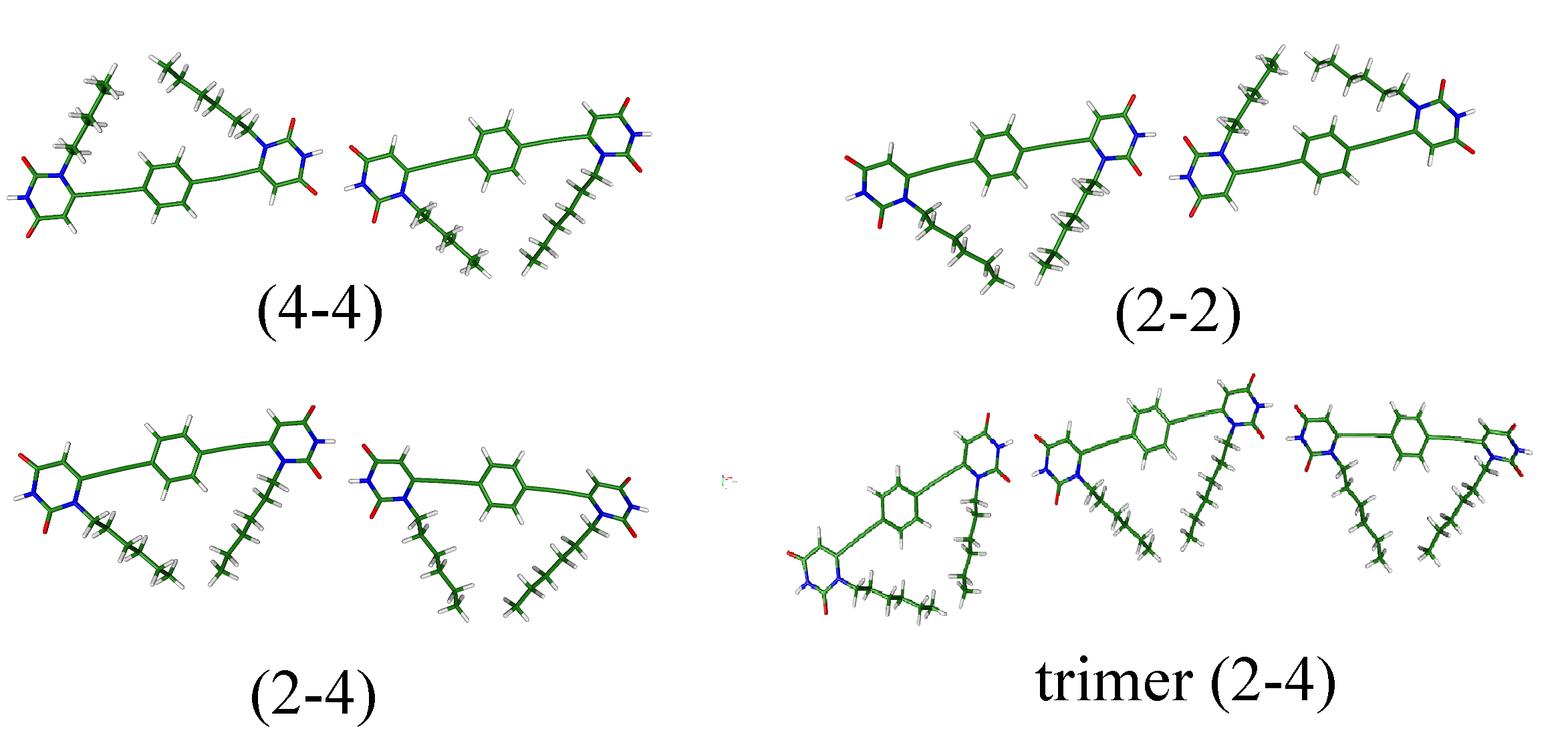}\\
  \caption{The investigated dimer and trimer configurations of {\MolTwo} in the theoretical study. For the dimer notation we indicate the labels of the carbon atoms (2 or 4) of the carbonyl groups taking part in the H-bond (according to Fig. \ref{fig:1}).}\label{fig:3}
\end{figure}

In Table \ref{tab:1} the calculated dimerization energies for
{\MolOne} and {\MolTwo} for three different configurations are presented (according to Fig. \ref{fig:3}). Looking at {\MolOne} first, the order of stability and the dimerization
energies obtained by the $\omega$B97X-D/6-311++G(3df,3pd)  method match previous theoretical results and experimental observations for {\MolOne}. Comparing  {\MolOne} and {\MolTwo} we observe the same tendency, the most stable dimer configuration being the (4-4) conformer followed by the (2-4)
and (2-2) conformers. The dimerization energy of {\MolTwo} indicates a slightly stronger coupling through H-bonds than for {\MolOne}, in agreement with the IR results presented in Fig. \ref{fig:2}.

\begin{table}[h]
\caption{Formation energies of the different isolated dimers for {\MolOne} and {\MolTwo}.}
\label{tab:1}
\begin{tabular}{ccc}
\hline
\textbf{Dimer} & \textbf{ \MolOne\ (kcal mol$^{-1}$)}  & \textbf{\MolTwo\ (kcal mol$^{-1}$)}\\
\hline
\textbf{(4-4)} & -13.3 & -14.8\\
\textbf{(2-2)} & -12.4 & -13.4\\
\textbf{(2-4)} & -12.6 & -13.9\\
\hline
\end{tabular}
\end{table}

\begin{figure}[h!]
  \includegraphics[width=15 cm]{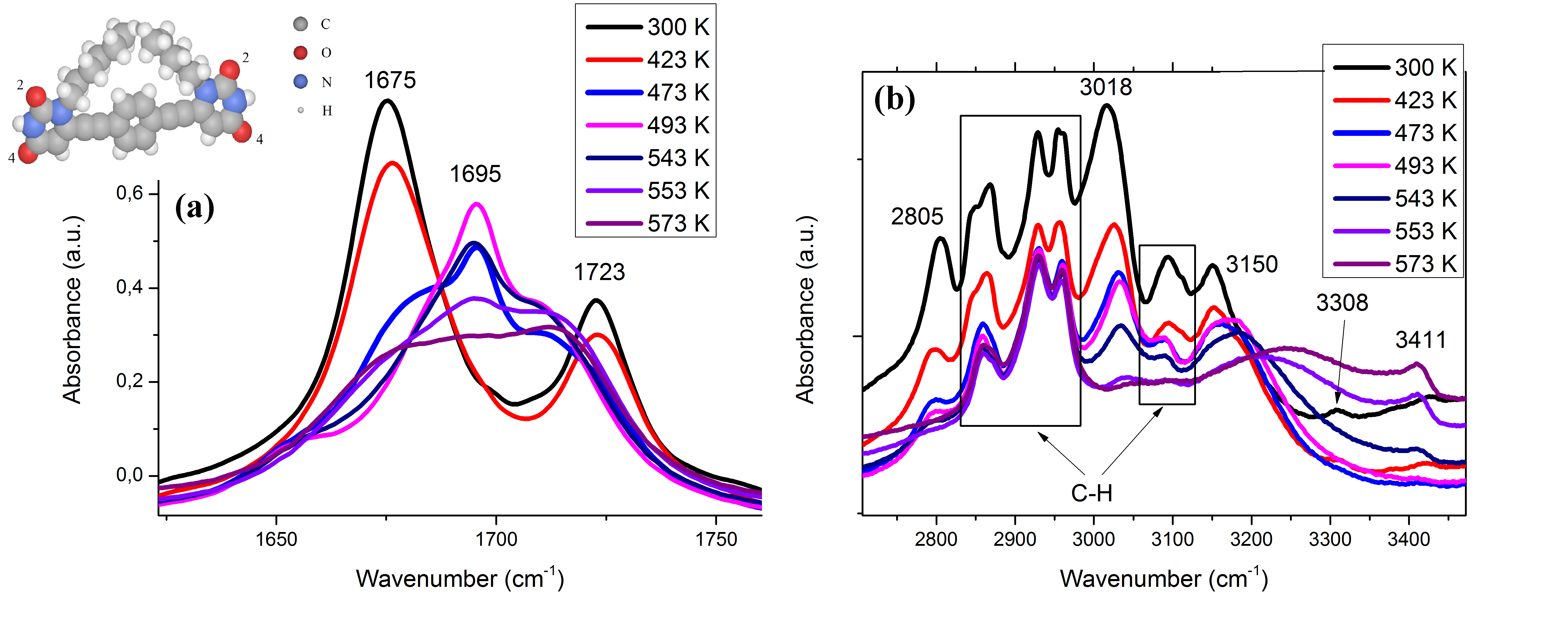}\\
  \caption{Experimental temperature-dependent IR spectra of {\MolTwo}. Above 473~K the band at 1675~{\cm-1} disappears and a new band appears at 1695~{\cm-1} which starts to decrease in intensity above 543~K. In the high frequency region the amine vibrations (3150~{\cm-1}) persist up to 543~K, above this temperature the free amine vibration band appears at 3411~{\cm-1} indicating the total melting of the H-bonds.
  }\label{fig:4}
\end{figure}

Figure \ref{fig:4} presents the temperature dependence of the infrared spectrum of {\MolTwo} from room temperature up to 573~K. There is an intermediate state between
the stable H-bonded phase at room temperature
and the case of melted H-bonds above 543~K.
Between 473~K and 543~K the intense C=O H-bonded band (at 1675~{\cm-1}) disappears and a new band appears at 1695~{\cm-1}. This temperature-induced transition can also be characterized by studying the behavior of the amine group. N-H
groups are always involved in H-bonding between molecules in this system. One-step melting of hydrogen bonds would result in the appearance of free N-H vibration bands; here, however, the H-bonded N-H band (at 3150~{\cm-1}) persists up to 543~K. All this indicates an intermediate phase, i.e. a phase transition at 473 K into a different, but still H-bonded, structure. The H-bonded N-H vibration, similarly to the H-bonded C=O stretch, shifts towards higher wavenumbers at 473~K. Above 543~K the H-bonded N-H band decreases in intensity and a new band appears at 3411~{\cm-1}, in the region characteristic of the free N-H vibrations.

\begin{figure}[h!]
\includegraphics[width=12 cm]{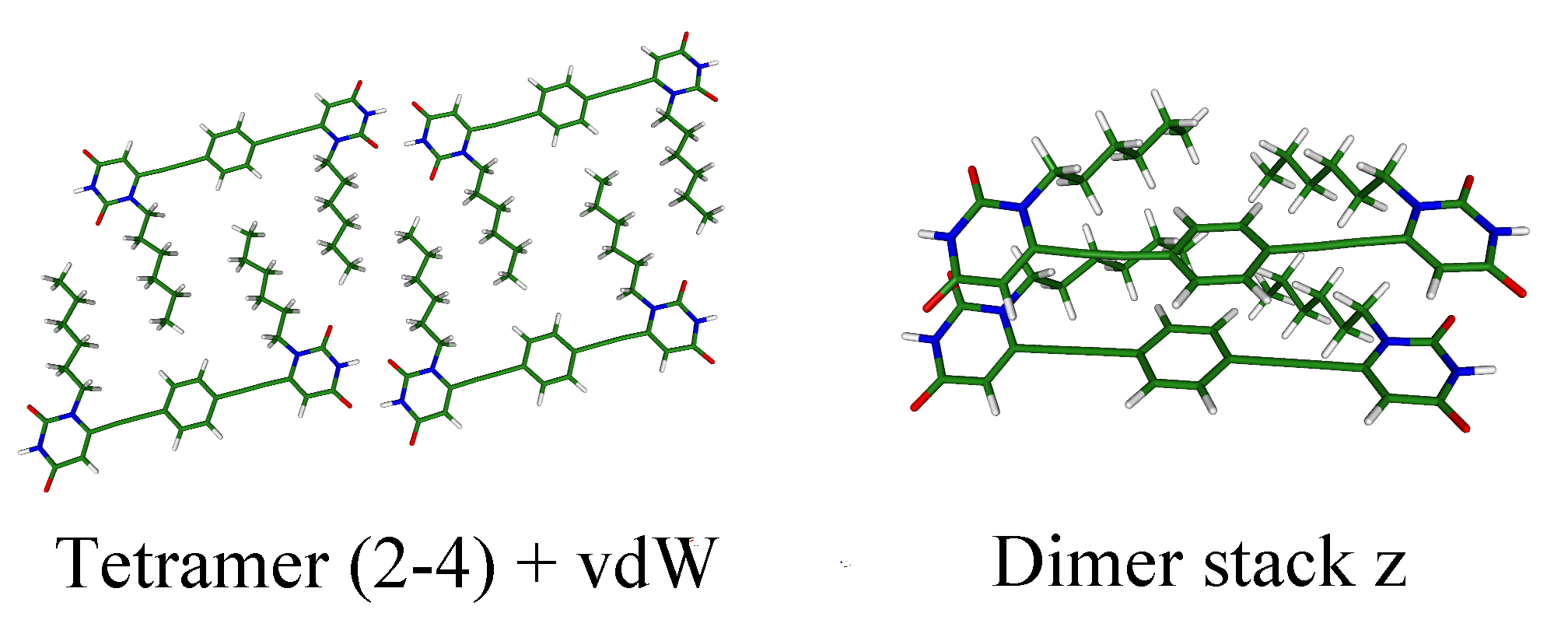}\\
\caption{Double-row wire structure organized mainly by double H-bonds and hexyl--hexyl London dispersion interactions between linear molecular assemblies (a).
Organization perpendicular to the molecular plane due to $\pi$--$\pi$ stacking of the aromatic rings (b).
  }\label{fig:5}
\end{figure}

In the following, we present a possible three dimensional arrangement of molecules {\MolTwo} following the experimental and theoretical data. We start from the structure organized on a surface, observed in a previous scanning tunneling microscopy (STM) study \cite{Llanes08}. This STM investigation, performed under ultrahigh vacuum on Ag(111) surfaces, provides important insight into the system
of noncovalent interactions that governs the self-assembly of molecule {\MolTwo}. The observed double-row supramolecular structure  is illustrated in Fig. \ref{fig:5}(a) for four units. Although the (4-4) dimer is found to be slightly more preferred energetically (cf. Table \ref{tab:1}),
the STM image verifies that a close-packed crystal structure can only be built with (2-4) type H-bonds. Furthermore, the all-cis configuration of the hexyl chains was found to be more abundant in Ref. \cite{Llanes08} even at submonolayer coverage in two dimensions. Therefore we continue with the theoretical study of the supramolecular structure shown in (Fig.  \ref{fig:5}(a)).

First of all, parallel linear assemblies are created by double hydrogen bonds between molecules {\MolTwo} (cf. Fig. \ref{fig:3}(d)). Theoretical investigation
reveals that double H-bonds can be formed independently on both ends
of molecule {\MolTwo}. Therefore this linear structure can grow to any length, since the stabilization energy, corresponding to the energy required for the extension of the chain by a single unit, is independent of chain size. This tendency is illustrated for the cases of trimers and tetramers in Table \ref{tab:2}.

\begin{table}[h!]
\caption{Formation energies for dimer, trimer and tetramer of the (2-4) conformer of {\MolTwo}.}
\label{tab:2}
\begin{tabular}{ccc}
\hline
\textbf{Dimer (kcal mol$^{-1}$)} & \textbf{Trimer (kcal mol$^{-1}$)} & \textbf{Tetramer (kcal mol$^{-1}$)}\\
\hline
-12.97 & -25.92 & -38.85 \\
 & ($\sim$ 2 x -12.97) & ($\sim$ 3 x -12.97)\\
\hline
\end{tabular}
\end{table}

The second organizing principle is the interdigitation
of the laterally placed hexyl chains stabilized by attractive dispersion interactions.
To approximate and compare the magnitude of the stabilization energies for the possible dimer arrangements,
single point DFT computations were carried out on the substructures of the tetramer in Fig. \ref{fig:5} (a).
Dimerization energies for the purely van der Waals  and H-bonded dimers were found to be -8.3~kcal$\cdot$mol$^{-1}$ and -14.6~kcal$\cdot$mol$^{-1}$, respectively.

Organization in the third direction, perpendicular to the aromatic planes, is governed by $\pi$-$\pi$ stacking
of the aromatic rings (Fig. \ref{fig:5} (b)). This is supported by previous investigations of $\pi$-$\pi$ interactions in uracil based crystals \cite{Czyznikowska07} and in stacked uracil dimers \cite{Pitonak08,Cooper08}. In case of molecule {\MolTwo}, $\pi$-$\pi$ stacking is particularly strong due to the three parallel aromatic rings. Dimerization energy for the dimer in Fig. \ref{fig:5} (b) is -33.6~kcal$\cdot$mol$^{-1}$, which is attributed  mainly to the $\pi$-$\pi$ stacking of the three aromatic rings and  partly to the attractive hexyl--hexyl dispersion interaction.

Since the interaction dominated by the dispersion of the hexyl groups turns out to be the lowest in energy, it is probable that it provides the "weakest link" to induce the transition at 473~K. Due to the large difference in the stabilization energies, the stronger double H-bonds are expected to  be relatively intact at that temperature. This argument is  supported by previous room-temperature AIMD IR computations on the (2-4)-type double H-bonded {\MolOne} dimer \cite{Szekrenyes12}. This structural change can also explain the shift of the H-bonded C=O band from 1675~{\cm-1} to 1695~{\cm-1}.
At lower temperature the additional attractive London dispersion of the hexyl groups, being parallel to the double H-bonds, shortens the H-bonds. This is illustrated in table \ref{tab:3} by comparing the computationally obtained geometrical parameters of the (2-4) dimer, the (2-4) linear tetramer and the double-row tetramer of Fig. \ref{fig:5} (a). Structures without the additional  stabilization of the hexyl--hexyl dispersion
have H-bond lengths around 1.825 \AA $\,$ and 1.815 \AA, while a significant decrease of cca. 0.06 \AA $\,$ can be found for the double-row structure. A shorter H-bond implies a lower force constant and thus lower frequency for the H-bonded C=O vibration, in accord with the experiment.

\begin{table}[h!]
\caption{Computed hydrogen bond length values for different complexes  of {\MolTwo}.}
\label{tab:3}
\begin{tabular}{ccc}
\hline
\textbf{Hydrogen bond length ({\AA})} & \textbf{C$_4$=\textbf{(O$\cdots$H)}-N} & \textbf{C$_2$=\textbf{(O$\cdots$H)}-N}\\
\hline
\textbf{dimer (2-4)} & 1.825 & 1.815\\
\textbf{tetramer (2-4), linear chain} & 1.822-1.825 & 1.816-1.818\\
\textbf{tetramer (2-4), double-row}  & 1.76, 1.78 & 1.73, 1.77\\
\hline
\end{tabular}
\end{table}

In light of all data, we assign the room-temperature
spectrum to a crystal structure of ordered double-row linear assemblies. These assemblies are formed via double H-bonds from monomers, and are organized into a double-row structure governed by dispersion interactions between laterally placed hexyl groups. Different layers of this practically planar arrangement are kept together by $\pi$-$\pi$ stacking of the aromatic rings.

\section {Conclusion}
\label{sec:conc}

The supramolecular ordering in the solid state of a bis-uracil based linear molecule has been studied using infrared spectroscopy and theoretical methods. Infrared spectra of the isolated monomers were obtained by the matrix isolation technique. Temperature dependence of the
vibrational bands affected by the H-bonds was followed in the solid state. Prior to the total
melting of the H-bonds an intermediate state was identified in the temperature profile. By analogy with previous STM measurements \cite{Llanes08} and supported by theoretical modeling, we suggest a structural phase transition at 473 K to explain this observation. The low-temperature structure in the three-dimensional solid is analogous to that on surfaces: double-row, ladder-like arrangements of {\MolTwo} units bound together by the $\pi$-$\pi$ stacking interactions of the aromatic rings perpendicular to the molecular plane. The two rows of molecules {\MolTwo} are held together mainly by attractive London dispersion interactions between the hexyl chains. This structure is consistent with temperature-dependent infrared and 2D STM measurements and theoretical results as well.

At an intermediate temperature of 473~K the H-bonded C=O band is shifted to a higher wavenumber, but the H-bonds are not disrupted yet, according to the frequency of the amine band.
The increase in the C=O frequency can be attributed to the disintegration of the weakest cohesive force, the London dispersion between the hexyl chains.
The full melting of the H-bonds was observed at 543~K, where the intensity of the H-bonded amine band decreases and a new band appears in the wavenumber region characteristic of free N-H vibrations.
We conclude that the supramolecular order in the three perpendicular directions of the solid crystal is governed by three different noncovalent interactions
characterized by their dominant components: double H-bonds, attractive dispersion between the hexyl groups, and $\pi$-$\pi$ stacking of the aromatic rings.
We find this arrangement  particularly interesting, since the strength of these interactions varies in a wide range, which
may permit direction-specific manipulations of this structure.

\section{Acknowledgment}

This research was funded by the Hungarian Scientific Research Fund (MI: grant no. OTKA K75877, IR measurements: grant No. OTKA NK105691) and supported by the supercomputing facility of the National Information Infrastructure Development Institute (NIIF). The work of PRN is supported by the UNKP-17-4-II New National Excellence Program of the Ministry of Human Capacities. The European Union and the European Social Fund have provided financial support to the project under the grant agreement no. T\'AMOP 4.2.1./B-09/KMR-2010-0003.

\end{document}